\documentclass[preprin,showpacs,preprintnumbers,eqsecnum,amsmath,amssymb,nofootinbib]{revtex4}
\usepackage{graphics,epsfig}
\usepackage{graphicx}
\usepackage{dcolumn}
\usepackage{bm}

\begin{document}
\baselineskip=0.8 cm
\title{ Three-dimensional charged Einstein-aether black holes and Smarr formula}

\author{Chikun Ding${}^{a,c}$ }
\email{Chikun_Ding@huhst.edu.cn; dingchikun@163.com}
\author{Changqing Liu${}^a$}
\author{ Anzhong Wang$^{b}$}
 \email{Anzhong_Wang@baylor.edu}
 \author{Jiliang Jing${}^c$}
 \email{jljing@hunnu.edu.cn}
  \affiliation{$^{a}$ Department of Physics, Hunan University of Humanities, Science and Technology, Loudi, Hunan
417000, P. R. China\\
$^{b}$ GCAP-CASPER, Physics Department, Baylor University, Waco, Texas 76798-7316, USA\\
$^{c}$Key Laboratory of Low Dimensional
Quantum Structures and Quantum Control of Ministry of Education,
and Synergetic Innovation Center for Quantum Effects and Applications,
Hunan Normal University, Changsha, Hunan 410081, P. R. China}

\vspace*{0.2cm}
\begin{abstract}
\baselineskip=0.6 cm
\begin{center}
{\bf Abstract}
\end{center}

We investigate behaviors of the three-dimensional  gravity coupled to a dynamical unit timelike vector: the aether, and present two new classes of exact charged solutions. When $c_{13}=0,\Lambda'=0$, we find the solutions is the usual BTZ black hole but now with an universal horizon.
 In the frame of black hole chemistry, we then calculate the temperature of the universal horizons and, construct the Smarr formulas and first law in the three cases: (quasi)asymptotically flat, aether asymptotically flat and quasi-BTZ black hole spacetime. We found these universal horizons obey an exact (or slightly modified) first law of black hole mechanics and may have an entropy and, black hole mass can be interpreted as enthalpy of spacetime. Then the  holography may be extended to these horizons under violating Lorentz symmetry.

\end{abstract}

 \maketitle

\vspace*{0.2cm}
\section{Introduction}

Lorentz invariance is one of the fundamental principles of Einstein's general relativity (GR) and modern physics.  However, Lorentz invariance may not be an exact symmetry at all energies  \cite{mattingly}. Any effective description must break down at a certain cutoff scale signaling the emergence of new physical degrees of freedom beyond that scale. For example, the hydrodynamics, Fermi's theory of beta decay \cite{bhattacharyya2} and quantization of GR \cite{shomer} at energies beyond the Planck energy. Lorentz invariance also leads to divergences in quantum field theory which can be cured with a short distance of cutoff that breaks it \cite{jacobson}. Astrophysical observations on high-energy cosmic rays seem to have proven this conjecture\cite{abdo}.
Einstein-aether theory can be considered as an effective description of Lorentz symmetry breaking in the gravity sector and has been extensively used in order to obtain quantitative constraints on Lorentz-violating gravity\cite{jacobson2}. And violations of Lorentz symmetry have
been used to construct modified-gravity theories that account
for Dark-Matter phenomenology without any actual Dark Mater \cite{bekenstein2004}.

 In Einstein-aether theory,  the Lorentz symmetry is broken only down to a rotation subgroup by the existence of a preferred time direction at every point of spacetime, i.e., existing a preferred frame of reference established by aether vector $u^a$.
This timelike unit vector field $u^a$ can be interpreted as a velocity four-vector of some medium substratum (aether, vacuum or dark fluid), bringing into consideration of non-uniformly moving continuous media and their interaction with other fields. Meanwhile, this theory can be
also considered as  a realization of dynamic self-interaction of complex systems moving with a spacetime dependant macroscopic velocity. As to an accelerated expansion of the universe, this dynamic self-interaction can produce the same cosmological effects as the dark energy \cite{balakin2}.

The introduction of the aether vector allows for some novel effects, e.g., matter fields can travel faster than the speed of light \cite{jacobson3}, new gravitational wave polarizations can spread at different speeds \cite{jacobson4}. Gravitational theories with breaking Lorentz invariance still allow the existence of black holes \cite{blas2,barausse,lin,UHs} which have an universal horizon. Instead of Killing horizon, it can trap excitations traveling at arbitrarily high velocities. So it attracts many concerns in many fields recently, e.g., Einstein-aether perfect fluid models\cite{latta}, its axionic extension\cite{balakin2016}, null aether theory\cite{gurses}, analogue black holes\cite{cropp2016}, etc.

Here we would fix our attention upon deriving a black hole solution of Einstein-aether theory. In 4-dimensions, the neutral black hole solutions\cite{berglund,bhattacharyya}, charged solutions\cite{ding}, slowly rotating solutions \cite{barausse2016} are found. On the other hand, there has been much interest in lower dimensional theories of gravity which there
is no dynamically propagating degrees of freedom: curvature is algebraically fixed by the matter content, in contrast to the four dimensional counterparts. The BTZ solution is the unique black hole of general relativity in 3-dimensions\cite{banados}. It generated a considerable amount of attention due to that it can be holographically described by a two-dimensional conformal field theory(AdS$_3/$CFT$_2$\cite{donnay}), whose foreseen applications in addressing conceptual issues of quantum gravity that become more tractable\cite{sotiriou}. So many lower dimensional solutions are found, such as, solutions in two-dimensions\cite{eling06}, numerical solutions of asymptotically Lifshitz spacetime\cite{basu} in three-dimension are found.

In another side, the relations between black hole physics and thermodynamics has been recognized for decades. Inspired by holography, novel studies of black holes with cosmological constant  provide some connections
between seemingly disparate theoretical concepts. Recently, black hole chemistry has emerged due to this thermodynamic connection\cite{frassino}. It tries to associate to each black
hole parameter a chemical equivalent in representations of
the first law. The present identifications are mass $M$ to thermal energy $E$, surface gravity $\kappa$ to  temperature $T$, horizon area $A$ to entropy $S$. However, one quantity---the pressure-volume term $PV$--- has no gravitational analogue. So the black hole chemistry  regards the cosmological constant $\Lambda$ of AdS spacetime as a thermodynamic pressure, and the mass $M$ as chemical enthalpy.

In this paper, we firstly  derive some new Lorentz-violating versions of the BTZ black hole in Einstein-aether theory. Nextly by using Komar integral method, seek for the Smarr relation and the first law in this black hole spacetime via the black hole chemistry.

The rest of the paper is organized as follows. In Sec. II we provide the background for the Einstein-Maxwell-aether theory studied in this paper. In Sec. III we review the first law and Smarr formula of BTZ black hole, and then show the procedure of how to construct a Smarr formula for spherically symmetric solutions. In Sec. IV, we first construct  two new classes of exact charged solutions, and then use them as examples to study the Smarr formula and  first law. In Sec. V, we present our main conclusions.

\section{Einstein-Maxwell-aether theory}

The general action for the Einstein-Maxwell-aether theory can be constructed by assuming that: (1) it is general covariant; and (2)  it  is a functional of only the spacetime metric $g_{ab}$, an unit timelike vector $u^a$ and Maxwell field $\mathcal{A}^a$, and involves  no more than two derivatives of them. So that the resulting field equations are second-order differential equations of  $g_{ab}$,   $u^a$ and $\mathcal{A}^a$. To simplify the problem, the couple between aether field and Maxwell one is ignored. Then,  the  Einstein-Maxwell-aether theory to be studied   in this paper is
 described by the  action,
\begin{eqnarray}
\mathcal{S}=
\int d^4x\sqrt{-g}\Big[\frac{1}{16\pi G_{\ae}}(\mathcal{R}-2\Lambda+\mathcal{L}_{\ae}+\mathcal{L}_M)\Big]\,. \label{action}
\end{eqnarray}

In terms of the tensor $Z^{ab}_{~~cd}$ defined as \cite{eling,garfinkle},
\begin{eqnarray}
Z^{ab}_{~~cd}=c_1g^{ab}g_{cd}+c_2\delta^a_{~c}\delta^b_{~d}
+c_3\delta^a_{~d}\delta^b_{~c}-c_4u^au^bg_{cd}\,,
\end{eqnarray}
 the aether Lagrangian $\mathcal{L}_{\ae}$ is given by
\begin{eqnarray}
-\mathcal{L}_{\ae}=Z^{ab}_{~~cd}(\nabla_au^c)(\nabla_bu^d)-\lambda(u^2+1),
\end{eqnarray}
where $c_i (i = 1, 2, 3, 4)$ are coupling constants of the theory.
The aether Lagrangian is therefore the sum of all possible terms for the aether field $u^a$ up to mass dimension two, and the constraint term $\lambda(u^2 + 1)$ with the Lagrange multiplier $\lambda$ implementing the normalization condition $u^2=-1$.
The source-free Maxwell Lagrangian $\mathcal{L}_M$ is given by
\begin{equation}
\mathcal{L}_M=-\frac{1}{4}\mathcal{F}_{ab}\mathcal{F}^{ab},
~\mathcal{F}_{ab}=\nabla_a\mathcal{A}_b-\nabla_b\mathcal{A}_a,
\end{equation}
where $\mathcal{A}_a$ is the electromagnetic potential four-vector.

The equations of motion, obtained by varying  the action (\ref{action}) with respect to $g_{ab}$, $u^a$, $\mathcal{A}^a$ and $\lambda$ are
\begin{eqnarray}\label{motion}
\mathcal{G}_{ab}+\Lambda g_{ab}=\mathcal{T}^{\ae}_{ab}+8\pi G_{\ae}\mathcal{T}^M_{ab},\quad {\AE}_a=0,\quad \nabla ^a\mathcal{F}_{ab}=0, \quad u^2=-1,
\end{eqnarray}
respectively, where the aether and Maxwell energy-momentum stress tensors $\mathcal{T}^{\ae}_{ab}$ and $\mathcal{T}^M_{ab}$ are given by
\begin{eqnarray}
\label{EMTs}
&&\mathcal{T}^{\ae}_{ab}=\lambda u_au_b+c_4a_aa_b-\frac{1}{2}g_{ab}Y^c_{~~d}\nabla_cu^d+\nabla_cX^c_{~~ab}
+c_1[(\nabla_au_c)(\nabla_bu^c)-(\nabla^cu_a)(\nabla_cu_b)],\nonumber \\
&&\mathcal{T}_{ab}^M
 =\frac{1}{16\pi G_{\ae}}\Big[-\frac{1}{4}g_{ab}\mathcal{F}_{mn}\mathcal{F}^{mn}
 +\mathcal{F}_{am}\mathcal{F}_{b}^{~m}\Big],
 \end{eqnarray}
with
\begin{eqnarray}
{\AE}_a=\nabla_bY^b_{~~a}+\lambda u_a+c_4(\nabla_au^b)a_b,\quad
Y^a_{~~b}=Z^{ac}_{~~~bd}\nabla_cu^d, \quad
 X^c_{~~ab}=Y^c_{~~(a}u_{b)}-u_{(a}Y^{~~c}_{b)}+u^cY_{(ab)} .
\end{eqnarray}
The acceleration vector $a^a$ appearing in the expression for the aether energy-momentum stress tensor is defined as the parallel transport of the aether field along itself, $a^a
\equiv \nabla_uu^a,$ where $\nabla_X\equiv X^b\nabla_b$.

Following \cite{berglund}, in spherically symmetry spacetime, the symmetry enforces $u^a$ hypersurface-orthogonal and becoming normal to one or more constant-radius hyperfurface that lies inside the Killing horizon. So one can let $\Sigma_U$ denote a surface orthogonal to the aether vector $u^a$, then $U$ is the aether time generated by $u^a$ that specifies each hypersurface in a foliation. As one moves in toward the origin, each $\Sigma_U$ hypersurface bends down to the infinite past, and asymptoting to a three-dimensional spacelike hypersurface on which $(u\cdot\chi)=0$, which implies that the Killing vector becomes tangent to $\Sigma_U$. This hypersurface is the universal horizon. Therefore, we are going to reduce these equations to a spherical symmetry case.

  We first define a set of basis vectors at every point in the spacetime,
so that we can project out various components of the equations of motion.
Let us first  take the aether field $u^a$ to be the basis vector. Then, pick up two spacelike unit vectors, denoted, respectively, by $m^a$ and $n^a$, both of which are normalized to unity,  mutually orthogonal, and lie on the tangent plane of the two-spheres $\mathcal{B}$ that foliate the hypersurface $\Sigma_U$. Finally, let us pick up $s^a$, a spacelike unit vector that is orthogonal to $u^a$, $m^a$, $n^a$, and  points ``outwards'' along a $\Sigma_U$ hypersurface, so we have the four tetrad, $
\left(u^a, s^a, m^a, n^a\right)$,  with the metric
\begin{eqnarray}
 ds^2 = -u_au_b + s_as_b + \hat{g}_{ab},
\end{eqnarray}
 where $\hat{g}_{ab}$ is
\begin{eqnarray}
 \hat{g}_{ab} =r^2(d\theta, d\phi)(d\theta, d\phi)\left(\begin{array}{cc}1&0\\0&\sin^2\theta\end{array}\right)=r^2d\theta^2+r^2\sin^2\theta d\phi^2,
\end{eqnarray}
 which are two-spheres $\mathcal{B}$ of 4D spherically spacetime. As for 3D case, we can let the ansatz of $u^a,s^a$ unchanged and now, $\mathcal{B}$ is one-spheres\footnote{ For $(n+2)$ dimensional spherically spacetime, the transverse space  could be a $n$-dimensional flat space. Then we can pick $n$ coordinates $y^i$ to describe this space, so that $\hat{g}_{ab}=b(r)q_{ij}(y)dy^idy^j$\cite{bhattacharyya2}. Any physical vector $X^a$ must be orthogonal to the transverse space, $\hat{g}_{ab}X^b=0$, hence we can let the ansatz of $u^a$ and $s^a$ unchanged and, all the following equations can be calculated directly from the 3D metric.}.

 By spherical symmetry, any physical vector $A^a$ has at most two non-vanishing components along, respectively,  $u^a$ and $s^a$, i.e., $A^a=A_1u^a+A_2s^a$.  In particular,
 the acceleration $a^a$  has only one component along $s^a$, namely,  $a^a=(a\cdot s)s^a$.
Similarly, any rank-two tensor $F_{ab}$ may have components along the directions of the bi-vectors $u_au_b,~u_{(a}s_{b)},~u_{[a}s_{b]},~s_as_b,~\hat{g}_{ab}$.
In the following, we study the expansion of the Maxwell field $\mathcal{F}^{ab}$,  Killing vector $\chi^a$,  surface gravity $\kappa$,  energy-momentum stress  tensors $\mathcal{T}_{ab}^{\ae}$ and $\mathcal{T}_{ab}^{M}$, and Ricci tensor $\mathcal{R}_{ab}$.
The given source-free Maxwell field $\mathcal{F}^{ab}$ can be formulated in
terms of four-vectors representing physical fields. They are
 the electric field $E^a$ and  magnetic excitation $B^a$ as\cite{balakin},
 \begin{eqnarray}
E^a=\mathcal{F}^{ab}u_b, \quad B^a=\frac{e^{abmn}}{2\sqrt{-g}}\mathcal{F}_{mn}u_b,
\end{eqnarray}
where $e^{abmn}$ is the Levi-Civita tensor. For source free Maxwell field or from Eq.(\ref{motion}), it can be shown
$B^a=0$, that is, there is no need to consider $e^{abmn}$. Then, we find
\begin{eqnarray}
\mathcal{F}^{ab}=-E^au^b+E^bu^a.
\end{eqnarray}
On the other hand,  the electric field is spacelike, since
$E^au_a=0$. So, we have $E^a=(E\cdot s)s^a$. Thus,  $\mathcal{F}_{ab}=(E\cdot s)\varepsilon^{II}_{ab}$, where $\varepsilon^{II}_{ab}=(-s_au_b+s_bu_a)$.

Using Gauss law, there has \cite{bhattacharyya2}
\begin{eqnarray}
\nabla^b[F(r)\varepsilon_{ab}^{II}]=0\quad\Longrightarrow\quad F(r)=\frac{F_0}{b(r)^{n/2}},
\end{eqnarray}
where $F_0$ is a constant, $b(r)=r^2$ for 4D or 3D spherically spacetime, $n$ is the number of dimension of sphere $\mathcal{B}$, i.e., $n$-sphere.
Here $\mathcal{B}$ is one-sphere for 3D spherically spacetime,
\begin{eqnarray}\label{3Dmetric}
ds^2=-e(r)dt^2+e^{-1}(r)dr^2+r^2d\phi^2,
\end{eqnarray}
 then $n=1$. From Eq. (\ref{motion}),
 we can see $(E\cdot s)=Q/r$, where $Q$ is an integral constant, representing the total charge of the space-time.
 Therefore, we have
\begin{eqnarray}
\label{Maxwellb}
\mathcal{F}_{ab}=\frac{Q}{r}(u_as_b-u_bs_a).
\end{eqnarray}

The Einstein,  aether  and Maxwell equations of motion (\ref{motion}) can be decomposed  by using the tetrad $u^a$, $s^a$ and $\hat{g}^{ab}$ defined above.
In particular,  the aether and electromagnetic energy-momentum stress tensors and the Ricci tensor can be cast, respectively, in the forms according directly to the 3D metric (\ref{3Dmetric}),
\begin{eqnarray}\label{projection}
&&\mathcal{T}^{\ae}_{ab}=\mathcal{T}^{\ae}_{uu}u_au_b-2\mathcal{T}^{\ae}_{us}u_{(a}s_{b)}
+\mathcal{T}^{\ae}_{ss}s_as_b+\hat{\mathcal{T}}_{\ae}\hat{g}_{ab},\nonumber\\
&&\mathcal{R}_{ab}=\mathcal{R}_{uu}u_au_b-2\mathcal{R}_{us}u_{(a}s_{b)}
+\mathcal{R}_{ss}s_as_b+\mathcal{\hat{R}}\hat{g}_{ab},\nonumber\\
&&\mathcal{T}^{M}_{ab}=\mathcal{T}^{M}_{uu}u_au_b-2\mathcal{T}^{M}_{us}u_{(a}s_{b)}
+\mathcal{T}^{M}_{ss}s_as_b+\hat{\mathcal{T}}_{M}\hat{g}_{ab}.
\end{eqnarray}
The coefficients of $\mathcal{T}^{\ae}_{ab}$ and $\mathcal{T}^{M}_{ab}$ in (\ref{projection}) can be computed from the general expression (\ref{EMTs}). The corresponding coefficients for $\mathcal{R}_{ab}$, on the other hand, are computed from the definition $[\nabla_a,~\nabla_b]X^c \equiv -\mathcal{R}^c_{~abd}X^d$ by choosing $X^a = u^a$ or $s^a$, and then contracting the resulting expressions again with $u^a$ and/or $s^a$ appropriately. The coefficients for the  three $(u, s)$ cross terms are
\begin{eqnarray}\label{usT}
\mathcal{T}^{\ae}_{us}=c_{14}\left[\hat{K}(a\cdot s)+\nabla_u(a\cdot s)\right],\quad
\mathcal{T}^{M}_{us}=0,\quad \mathcal{R}_{us}=(K_0-\hat{K})\hat{k}-\nabla_s\hat{K},
\end{eqnarray}
where
\begin{eqnarray}\label{usTa}
\nabla_{[a}s_{b]} \equiv - K_0 u_{[a}s_{b]},\;\;\;
\hat{k} \equiv \frac{1}{2} g^{ab} {\cal{L}}_s\hat g_{ab},\;\;\;
\hat{K} \equiv \frac{1}{2} g^{ab} {\cal{L}}_u\hat g_{ab},
\end{eqnarray}
with $K \; (\equiv K_0+\hat{K}) $ being  the trace of the extrinsic curvature of the  hypersurface $\Sigma_U$.
The aether equation $s\cdot{\AE}=0$ and   the $us$-component  $\mathcal{R}_{us}=\mathcal{T}^{\ae}_{us} + 8\pi G_{\ae} \mathcal{T}^{M}_{us}$ yield
\begin{eqnarray}\label{usA}
 && c_{123}\nabla_s K_0-(1-c_{13})(K_0-\hat{K})\hat{k}+(1+c_2)\nabla_s\hat{K}=0,\\
 \label{usB}
&&  c_{123}\nabla_s K-(1-c_{13})\mathcal{T}^{\ae}_{us} = 0.
\end{eqnarray}

After rewriting the motion Eq. (\ref{motion}) as
\begin{eqnarray}\label{motion2}
\mathcal{R}_{ab}-2\Lambda g_{ab}=\mathcal{T}^{\ae}_{ab}-g_{ab}\mathcal{T}^{\ae}+8\pi G_{\ae}\big[\mathcal{T}^M_{ab}-g_{ab}\mathcal{T}^M\big],
\end{eqnarray}
where $T=g^{ab}T_{ab}$, then the  $uu$-, $ss$- and $\hat{g}$-components of the gravitational field equations  give
\begin{eqnarray}\label{uussA}
\nabla\cdot a-(1-c_{13})(K_0^2+\hat{K}^2)+2\Lambda-c_{123}\nabla_c(Ku^c)-(1+c_2)\nabla_uK=0,\\
\label{uussB}
\nabla_c\big[(K_0+c_{13}\hat{K}+c_2K)u^c\big]-\big[a^2+\hat{k}^2+\nabla_s(a\cdot s+\hat{k})+2\Lambda\big]-c_{14}(\nabla\cdot a-a^2)=0,\\\label{uussC}
\mathcal{R}+(1-c_{14})\nabla\cdot a+[a^2+\hat{k}^2+
\nabla_s(a\cdot s+\hat{k})]\nonumber\\
-\nabla_c[\big((1-c_{13})K_0-c_2K\big)u^c]-\nabla_uK-(K_0^2+\hat{K}^2)
-2\Lambda-\frac{Q^2}{2r^2}=0,
\end{eqnarray}
where $\mathcal{R}=g^{ab}\mathcal{R}_{ab}=-e''(r)-2e'(r)/r$. In the next sections, we will use these equations to obtain new  black holes solutions.

\section{Smarr formula}

In this section we firstly review the Smarr formula and first law of BTZ black hole spacetime via black hole chemistry. And then using Komar integral method, we show the procedure of deriving the Smarr formula in 3-dimensional Einstein-Maxwell-aether theory.

In three dimensions, the static BTZ black hole reads\cite{banados}
\begin{eqnarray}\label{btz}
ds^2=-fdt^2+\frac{dr^2}{f}+r^2d\phi^2,\;f=-m+r^2/l^2,
\end{eqnarray}
where $1/l^2=-\Lambda$. The first law of black hole mechanics and the Smarr formula are\cite{kubiznak}
\begin{eqnarray}\label{btzsmarr}
dM=TdS+VdP,~~~~0=TS-2PV,
\end{eqnarray}
where the temperature $T=f'(r_+)/4\pi=r_+/2\pi l^2$, the entropy $S=A/4=\pi r_+/2$, the black hole mass $M=m/8=r_+^2/8l^2$ and, $P=1/8\pi l^2$ is thermodynamic pressure, its thermodynamic conjugate volume $V=\partial M/\partial P|_S=\pi r_+^2$, $r_+$ is the horizon radius. The pressure-volume term reinterprets $M$ as the enthalpy of the black hole\cite{kastor}: the
energy required to both form a black hole and place it into
its cosmological environment. In the next section, these exact or slightly modified first law and Smarr formulas will be constructed for those Lorentz-violating BTZ-like black holes by the following procedure.

Now  we shall present the process of deriving Smarr formulas of the universal horizons for general $2+1$  static and spherically symmetric Einstein-Maxwell-aether black holes. Let us first consider the geometric identity \cite{BCH},
\begin{eqnarray}
\label{ID}
\mathcal{R}_{ab}\chi^b=\nabla^b(\nabla_a\chi_b).
\end{eqnarray}
The derivative of the Killing vector $\chi^a=-(u\cdot\chi)u^a+(s\cdot\chi)s^a$ is given by
\begin{eqnarray}\label{kappa0}
\nabla^a\chi^b=-2\kappa u^{[a}s^{b]},\quad
\end{eqnarray}
where $\kappa$ denotes  the surface gravity usually defined in GR, and is given by
\begin{eqnarray}\label{kappa1}
\kappa =\sqrt{-\frac{1}{2}(\nabla_a\chi_b)(\nabla^a\chi^b)}=-(a\cdot s)(u\cdot\chi)+K_0(s\cdot\chi).
\end{eqnarray}

From the Einstein field equations (\ref{motion2}), we find that
\begin{eqnarray}
\label{IDa}
&&2\Lambda g_{ab}\chi^b=\nabla^b\Big(\Lambda r\; u_{[a}s_{b]}\Big),\nonumber\\
&& 8\pi G_{\ae} (\mathcal{T}^M_{ab}-g_{ab}\mathcal{T}^M)\chi^b =0,\nonumber\\
&&   \left(\mathcal{T}^{\ae}_{ab} -  g_{ab}\mathcal{T}^{\ae}\right)\chi^b = \nabla^b\big[\left(c_{123} K  - c_{13} K_0\right) \left(s\cdot \chi\right)u_{[a}s_{b]}\big] .
\end{eqnarray}
Then Eq.(\ref{ID}) can be cast in the form,
\begin{eqnarray}
\label{SmarrFa}
\nabla_bF^{ab}=0,\;\;\; F^{ab} \equiv 2q(r)u^{[a}s^{b]},
\end{eqnarray}
where
\begin{eqnarray}
\label{SmarrFb}
&& q(r)  \equiv\Lambda r-(a\cdot s)(u\cdot \chi)+\left[(1-c_{13})K_0 +c_{123}K\right](s\cdot \chi).
\end{eqnarray}
On the other hand, comparing Eq.(\ref{SmarrFa}) with the soucre-free Maxwell equations  (\ref{motion}), we find that its solution must also take the form (\ref{Maxwellb}), that is,
$q(r) = q_0/r$.
On the other hand, using Gauss' law,  from Eq.(\ref{SmarrFa}) we find that
\begin{eqnarray}\label{tmassB}
   0 = \int_{\Sigma}{\left(\nabla_bF^{ab}\right)   d\Sigma_a} =   \int_{{\cal{B}}_{\infty}}{F^{ab}   d\Sigma_{ab}} - \int_{{\cal{B}}_{H}}{F^{ab}   d\Sigma_{ab}}
   =   \int_{{\cal{B}}_{\infty}}{qdA} - \int_{{\cal{B}}_{H}}{q dA}.
\end{eqnarray}
Here $d\Sigma_a$ is the surface element of a spacelike  hypersurface $\Sigma$, and $dA=rd\phi$ is the area element of 1-dimensional sphere. The boundary $\partial\Sigma$ of $\Sigma$
consists of the boundary  at spatial infinity ${\cal{B}}_{\infty}$,  and the horizon ${\cal{B}}_{H}$, either the Killing or the universal.
Note that Eq.(\ref{tmassB}) is nothing but the conservation law of the flux of $F^{ab}$. Comparing the above expression   and
Eq.(\ref{SmarrFb}), we find the following Smarr formula in 3-dimensions\cite{kastor},
\begin{eqnarray}
0\cdot MG_{\ae}=\frac{q_{H}A_{H}-q_{\infty}A_{\infty}}{8\pi},
\end{eqnarray}
where $A\infty$ and $A_{H}$ are  the area of infinite boundary and the universal or Killing horizon, however now $M$ is the enthalpy instead of the total energy of spacetime defined in the asymptotic aether rest frame.
 The $q_\infty$ and $q_H$ are the value of $q$ in (\ref{SmarrFb}) at the infinite and universal horizon or Killing horizon.

In GR, from Eq. (\ref{kappa0}) and (\ref{ID}), the $q_H$ is just the surface gravity $\kappa_H$ on the usual Killing horizon. Now in the presence of aether field, it contains the aether contribution and becomes complicated.
However, the first law for the aether black hole may still be obtained via a variation of these Smarr relations.
In the next section we consider it for  two new classes of exact charged aether black hole solutions.

For  the surface gravity at the universal horizon, when one  considers the peeling behavior of particles moving at any speed, i.e., capturing the role of the aether in the propagation of the physical rays, one finds that the  surface gravity at the universal horizon  is \cite{cropp,jacobson5,lin}
\begin{eqnarray}\label{kappa}
\kappa_{UH}\equiv \frac{1}{2}\nabla_u(u\cdot\chi)
= \left.\frac{1}{2}\left(a\cdot s\right)\left(s\cdot \chi\right) \right|_{r= r_{UH}},
\end{eqnarray}
where in the last step we used the fact that $\chi_a$ is a Killing vector,  $\nabla_{(a}\chi_{b)} =0$.  It must be noted that this is different from the surface gravity defined in GR  by Eq.(\ref{kappa1}). In particular, at the universal horizon we have $u\cdot \chi = 0$, and Eq.(\ref{kappa1}) yields,
\begin{eqnarray}\label{kappa2}
\kappa\left(r_{UH}\right) = \left. K_0(s\cdot\chi)\right|_{r= r_{UH}}.
\end{eqnarray}

\section{Exact Solutions of charged aether black holes}

To construct exact solutions of charged aether black holes, let us first choose the Eddington-Finklestein coordinate system, in which
the    metric takes the form
\begin{eqnarray}\label{metric}
ds^2=-e(r)dv^2+2dvdr+r^2d\phi^2,
\end{eqnarray}
and the corresponding timelike Killing  and aether vectors are
\begin{equation}
  \chi^a=(1,0,0),\quad u^a=\big(\alpha,~\beta,~0,\big),\quad u_adx^a=\big(-e\alpha+\beta,~\alpha,~0,\big)\left(
\begin{array}{c}
 dv\\
 dr\\  d\phi
\end{array}
\right),
\end{equation}
where $\alpha(r)$ and $\beta(r)$ are  functions of $r$ only. Then,  the metric can be written as  $g_{ab}=-u_au_b+s_as_b+\hat{g}_{ab}$, where we have the
 constraints $u^2=-1, ~s^2=1,~ u\cdot s=0$.

Some quantities that explicitly appear in Eqs.(\ref{usA})-(\ref{uussC}) are  \cite{bhattacharyya}
\begin{eqnarray}
\label{AS}
 (a\cdot s)=-(u\cdot \chi)',\quad K_0=-(s\cdot \chi)',\quad\hat{K}=-\frac{(s\cdot \chi)}{r},\quad\hat{k}=-\frac{(u\cdot \chi)}{r},
\end{eqnarray}
where a prime $(')$ denotes a derivative with respect to $r$. And $\alpha(r),~\beta(r)$ and $ e(r)$ are
\begin{eqnarray}\label{abe}
 \alpha(r)=\frac{1}{(s\cdot\chi)-(u\cdot\chi)},\quad \beta(r)=-(s\cdot \chi), \quad e(r)=(u\cdot\chi)^2-(s\cdot\chi)^2.
\end{eqnarray}
Then, from Eqs.(\ref{kappa}) and (\ref{kappa2}) we obtain
\begin{eqnarray}\label{abeA}
 \kappa_{UH} =-\left. \frac{1}{2}\left(u\cdot \chi\right)' (s\cdot\chi)\right|_{UH},\quad
  \kappa(r_{UH}) =-\left. \left(s\cdot \chi\right)' (s\cdot\chi)\right|_{UH}.
\end{eqnarray}
Clearly, in general $ \kappa_{UH} \not= \kappa(r_{UH})$.

Substituting Eq.(\ref{AS}) into (\ref{usT}), a straightforward  calculation yields
 \begin{eqnarray}
 \mathcal{R}_{us}=0.
\end{eqnarray}
So that the $us$-component gravitational field motion equation  $\mathcal{R}_{us}=\mathcal{T}^{\ae}_{us} + 8\pi G_{\ae} \mathcal{T}^{M}_{us}$ yields
\begin{eqnarray}
\mathcal{T}^{\ae}_{us}=0,
\end{eqnarray}
and the aether field motion Eq.(\ref{usB}) gives
\begin{eqnarray}
c_{123}\nabla_sK=0.
\end{eqnarray}
They both together with Eqs. (\ref{usT}) and (\ref{AS}) lastly give
\begin{eqnarray}\label{aetherEq}
&&\mathcal{T}^{\ae}_{us}=c_{14}\frac{(s\cdot\chi)}{r^2}[r(u\cdot\chi)']'=0,
\nonumber\\
&&c_{123}\nabla_sK=c_{123}(u\cdot\chi)\big[\frac{1}{r}\big(r(s\cdot\chi)\big)'
\big]'=0.
\end{eqnarray}
It is easy to see that there are many ways for satisfying these two equations, in the following,  we shall consider only two special cases $c_{14}=0,~c_{123}\neq0$ and $c_{123}=0,~c_{14}\neq0$ to obtain both classes of exact solutions.

\subsection{Exact solutions for $c_{14}=0$}

When the coupling constant $c_{14}$ is set to zero and $c_{123}\neq 0$, from Eqs.(\ref{usB}) and (\ref{aetherEq}) one can see the quantity $\nabla_sK$ has to be vanished, i.e., $\nabla_sK=0$. So, the trace of the extrinsic curvature $K$ of the $\Sigma_U$ hypersurface is a  constant and, Eq. (\ref{aetherEq}) gives
 \begin{eqnarray}\label{sx}
 &&(s\cdot\chi)=\Lambda'r+\frac{r_{\ae}}{r},
\end{eqnarray}
where $\Lambda',r_{\ae}$ are some constants.
Since a vanishing $(s\cdot\chi)$ signifies of the aether $u^a$ aligning with Killing vector $\chi^a$, so both constants measure the misalignment of the aether. When $\Lambda'=0$, then the aether aligns Killing vector at infinite, say, aether asymptotically flat.
   Substituting Eq.(\ref{sx}) into  (\ref{uussC}), we obtain
   \begin{eqnarray}\label{aether}
 &&(u\cdot\chi)=-\sqrt{-m+2\Lambda'r_{\ae}+\bar\Lambda r^2-\frac{Q^2}{2}\ln\frac{r}{l_{eff}}+(1-c_{13})\frac{r^2_{\ae}}{r^2}}\;,\\
 \label{metric1}&&e(r)=-m+(\bar\Lambda-\Lambda'^2) r^2-\frac{Q^2}{2}\ln\frac{r}{l_{eff}}-\frac{c_{13}r^2_{\ae}}{r^2},
\end{eqnarray}
where $m$ is an integral constant which then shows the mass of the black hole, $1/l_{eff}=\sqrt{\bar\Lambda}$ and, $\bar\Lambda=-\Lambda+(1+c_{123}+c_2)\Lambda'^2$.
 It is BTZ-like black hole, and when $Q=0$, it can reduce to the static solution given  in Ref.  \cite{sotiriou} via Ho\v{r}ava-Lifshitz gravity.
 If $c_{13}=0$ and $\Lambda'=0$, it is the usual BTZ black hole, but now with an universal horizon, which is very interesting.

The metric (\ref{metric1}) can be (quasi) asymptotically dS, flat or AdS, when $(\bar\Lambda-\Lambda'^2)<0,=0$, or $>0$, respectively. And there is a condition that $(u\cdot\chi)^2\geqslant0$. From (\ref{aether}) and at large $r$, the term $\bar\Lambda r^2$ dominates $(u\cdot\chi)^2$; at little $r$, the term $(1-c_{13})$ dominates, therefore  $\bar\Lambda$ should be positive, $\bar\Lambda>0$ and $c_{13}<1$. Hence for AdS asymptotics, it is $(c_2+c_{123})\Lambda'^2>\Lambda$ and $m>0$; for quasi-flat, it is $(c_2+c_{123})\Lambda'^2=\Lambda$ and $m<0$($m=-1$ for flat); for dS, it is $(c_2+c_{123})\Lambda'^2<\Lambda<(1+c_2+c_{123})\Lambda'^2$ and $m<0$. If $\Lambda'=0$, there are also three asymptotics,  dS, flat or AdS for $\Lambda>0,=0$ or $<0$, respectively.

The location of the universal horizon $r_{UH}$ is the largest root of equation $u\cdot\chi=0$. Meanwhile, $u\cdot\chi$ is a physical component of the aether, and should be regular and real everywhere. However,  from Eq.(\ref{aether}) one can see that in the region $r_{-}<r<r_{UH}$, this term becomes purely imaginary,
where $r_{-}$ is another root of $u\cdot\chi=0$, unless the two real roots coincide. Then, $r_{\ae}$ becomes a function of $m$. That is, the global existence of the
aether reduces the number of four independent constants ($m,\Lambda', r_{\ae}, Q$) to three,  ($m,\Lambda', Q$).  Thus,  from  $(u\cdot\chi)^2 = 0$ and $d(u\cdot\chi)^2/dr = 0$  \cite{lin}, we find
\begin{eqnarray}\label{ruh1}
r^2_{\ae}=\frac{1}{1-c_{13}}\left(\bar\Lambda r_{UH}^4
-\frac{Q^2}{4}r_{UH}^2\right),\quad
 m=2\bar\Lambda r_{UH}^2-\frac{Q^2}{2}\Big(\ln\frac{r_{UH}}{l}+
\frac{1}{2}\Big)+2\Lambda'r_{\ae}.
\end{eqnarray}

Note that from (\ref{ruh1}), if the black hole and the aether both asymptotically flat, i.e., $\Lambda'=0,\bar\Lambda=0$, there is no universal horizon which has been showed in Ref.\cite{sotiriou}. However, if $r_{\ae}=0$, there still exists an universal horizon due to presence of electric charge $Q$, and we will call it quasi-BTZ black hole.

Now let us derive the Smarr formula and the first law at the universal horizon.  The surface gravity at the universal horizon can be computed via (\ref{kappa}) and given by
\begin{eqnarray}\label{kappauh1}
 \kappa_{UH}=\frac{1}{2}\nabla_u(u\cdot\chi)|_{r_{UH}}=\frac{1}{2r_{UH}}
\sqrt{\bar\Lambda r_{UH}^2-\frac{Q^2}{8}}\left[
 \sqrt{\frac{4\bar\Lambda r_{UH}^2-Q^2}{1-c_{13}}}+2\Lambda'r_{UH}\right].
\end{eqnarray}
If one uses definition (\ref{kappa2}), then this surface gravity is
\begin{eqnarray}\label{kappauh11}
 \kappa(r_{UH})=K_0(s\cdot\chi)|_{r_{UH}}=-\Lambda'^2r_{UH}
 +\frac{1}{(1-c_{13})r_{UH}}
\left(\bar\Lambda r_{UH}^2-\frac{Q^2}{4}\right),
\end{eqnarray}
which is different.

From (\ref{SmarrFb}) and (\ref{tmassB}), one get the Komar integral relation
\begin{eqnarray}\label{smarr1}
0\cdot MG_{\ae}=2\pi r_{UH}[\Lambda r_{UH}+(1-c_{13})K_0(s\cdot\chi)+c_{123}K(s\cdot\chi)]+4\pi c_{123}\Lambda'r_{\ae}+\frac{\pi Q^2}{2}.
\end{eqnarray}
Then by using surface gravity (\ref{kappauh1}) and the knowledge of black hole chemistry, we can derive the Smarr formula and first law of these Lorentz-violating black holes.
In the following, we study on these thermodynamic properties of the universal horizons in three cases.

\subsubsection{(Quasi)asymptotically flat black hole}

Consider firstly uncharged and quasi-asymptotically flat spacetime, i.e., $Q=0$ and $\Lambda'^2=\Lambda/(c_2+c_{123})$. From (\ref{metric1}), this metric is
\begin{eqnarray}e(r)=-m-\frac{c_{13}r^2_{\ae}}{r^2}.
 \end{eqnarray}
To ensure there exists a black hole, it requires that $m<0$ when $0<c_{13}<1$; $m>0$ when $c_{13}<0$. If $m=-1$, it is exact asymptotically flat spacetime, therefore these black holes are termed (quasi)asymptotically flat. From (\ref{ruh1}), then the mass $m$ is
\begin{eqnarray}
 m=\left\{
\begin{array}{c}
  2\bar\Lambda r_{UH}^2\left(1-\frac{1}{\sqrt{1-c_{13}}}\right),\quad (0<c_{13}<1),
 \\ 2\bar\Lambda r_{UH}^2\left(1+\frac{1}{\sqrt{1-c_{13}}}\right),\quad (c_{13}<0).
\end{array}
\right.
\end{eqnarray}
In the following, we let $m>0$.

 Dividing both sides of Eq. (\ref{smarr1}) by $8\pi c_{13}\sqrt{1-c_{13}}/(1+\sqrt{1-c_{13}})$, we can obtain the Smarr formula
\begin{eqnarray}
0=TS-2PV,
\end{eqnarray}
where
\begin{eqnarray}
&&T=\frac{\kappa_{UH}}{2\pi}=\Big[1+\frac{1}{\sqrt{1-c_{13}}}\Big]\frac{r_{UH}}{2\pi l_{eff}^2},\quad S=A/4=\frac{\pi r_{UH}}{2},\quad
M=\frac{m}{16}=\Big[1+\frac{1}{\sqrt{1-c_{13}}}\Big]\frac{r_{UH}^2}{8l_{eff}^2},\nonumber\\
&&P=\frac{1}{8\pi l_{eff}^2},\quad V=\frac{\partial M}{\partial P}\Big|_S=\Big[1+\frac{1}{\sqrt{1-c_{13}}}\Big]\pi r_{UH}^2,\quad(\frac{1}{l_{eff}}=\sqrt{\bar\Lambda}).
\end{eqnarray}
Based on these definitions, it is straightforward to verify the first law
\begin{eqnarray}
dM=TdS+VdP,
\end{eqnarray}which is the same as (\ref{btzsmarr}).

\subsubsection{Aether asymptotically flat black hole}
Secondly the uncharged black hole and the aether aligning with Killing vector at infinite, i.e., $Q=0$ and $\Lambda'=0$.
From (\ref{metric1}), this metric is
\begin{eqnarray}e(r)=-m-\Lambda r^2-\frac{c_{13}r^2_{\ae}}{r^2}.
 \end{eqnarray}
There are both Killing horizons
\begin{eqnarray}r^2_{\pm}=-\frac{m}{2\Lambda}
\left[1\pm\sqrt{1-\frac{\Lambda}{m^2}c_{13}r^2_{\ae}}\right],
 \end{eqnarray}
and have three asymptotics, i.e., AdS when $(\Lambda<0,m>0,c_{13}<0)$; dS when   $(\Lambda>0,m<0,0<c_{13}<1)$; (quasi)flat when $\Lambda=0$.

 Dividing both sides of Eq. (\ref{smarr1}) by $8\pi$, we can obtain the slightly modified Smarr formula
\begin{eqnarray}
0=\sqrt{1-c_{13}}TS-2PV,
\end{eqnarray}
where
\begin{eqnarray}
&&T=\frac{\kappa_{UH}}{2\pi}=\frac{1}{\sqrt{1-c_{13}}}\frac{r_{UH}}{2\pi l^2},\quad S=A/4=\frac{\pi r_{UH}}{2},\quad
M=\frac{m}{16}=\frac{r_{UH}^2}{8l^2},\nonumber\\
&&P=\frac{1}{8\pi l^2},\quad V=\frac{\partial M}{\partial P}\Big|_S=\pi r_{UH}^2,\quad(\frac{1}{l}=\sqrt{|\Lambda|}).
\end{eqnarray}
Based on these definitions, it is straightforward to verify the slightly modified first law
\begin{eqnarray}
dM=\sqrt{1-c_{13}}TdS+VdP.
\end{eqnarray}
\subsubsection{Quasi-BTZ black hole}
Thirdly, the case of $r_{\ae}=0$.
From (\ref{metric1}), this metric is
\begin{eqnarray}\label{quasibtz}
e(r)=-m+(\bar\Lambda-\Lambda'^2) r^2-\frac{Q^2}{2}\ln\frac{r}{l_{eff}},
 \end{eqnarray}
which is only different cosmological constant from BTZ metric \cite{banados}, so termed quasi-BTZ black hole. Its universal horizon is
\begin{eqnarray}r_{UH}=\frac{Q}{2\sqrt{\bar\Lambda}}.
 \end{eqnarray}

 Dividing both sides of Eq. (\ref{smarr1}) by $8\pi$, we can obtain the slightly modified Smarr formula
\begin{eqnarray}
0=-\frac{\sqrt{2}}{\Lambda'l_{eff}}TS-2PV,
\end{eqnarray}
where
\begin{eqnarray}
&&T=\frac{\kappa_{UH}}{2\pi}=\frac{\Lambda'}{2\sqrt{2}\pi}\frac{r_{UH}}{ l_{eff}},\quad S=A/4=\frac{\pi r_{UH}}{2},\quad
M=\frac{m}{8}=\frac{Q^2}{16}(\frac{1}{2}-\ln\frac{r_{UH}}{l_{eff}}),\nonumber\\
&&P=\frac{1}{8\pi l_{eff}^2},\quad V=\frac{\partial M}{\partial P}\Big|_{S,Q}=-\frac{\pi Q^2}{4} l_{eff}^2,\quad \Phi=\frac{\partial M}{\partial Q}\Big|_{S,P}=\frac{Q}{8}(\frac{1}{2}-\ln\frac{r_{UH}}{l_{eff}}),
\end{eqnarray}
where $\frac{1}{l_{eff}}=\sqrt{\bar\Lambda}$.
Based on these definitions, it is straightforward to verify the slightly modified first law
\begin{eqnarray}
dM=-\frac{\sqrt{2}}{\Lambda'l_{eff}}TdS+VdP+\Phi dQ.
\end{eqnarray}

In summary, for these three cases, the exact or the slightly modified first law of black hole mechanics can be construct, which show the corresponding universal horizons would have a thermodynamic interpretation.
For the general cases of nonzero $Q,\Lambda',r_{\ae}$, and $\Lambda'^2\neq\Lambda/(c_2+c_{123})$, instead of above three cases, one cannot build the Smarr formula and first law. As for causes, there may be the black hole mass need to be modified, see Ref.\cite{ding,ding2} for 4-dimensions, or couples between the charge $Q$ and aether field should be considered together\cite{balakin,balakin2016}. These open issues are our future works.

\subsection{Exact solutions for $c_{123}=0$}

When the coupling constant $c_{123}$ is set to zero and $c_{14}\neq 0$,   Eq. (\ref{aetherEq}) gives
 \begin{eqnarray}\label{ux}
 &&(u\cdot\chi)=u_0\ln\frac{r}{ r_{UH}},
\end{eqnarray}
where $u_0$ is  a constant and $r_{UH}$ shows the location of universal horizon.
   Substituting it into  Eqs.(\ref{uussC}), we obtain
\begin{eqnarray}
 &&(s\cdot\chi)=\frac{1}{\sqrt{1-c_{13}}}\sqrt{(1-c_{13})m+\Lambda r^2+(c_{14}u_0^2+\frac{Q^2}{2})\ln\frac{r}{l}+u_0^2\ln^2\frac{r}{r_{UH}}}\;,\\
 &&e(r)=-m-\frac{1}{1-c_{13}}\Big[\Lambda r^2+(c_{14}u_0^2+\frac{Q^2}{2})\ln\frac{r}{l}+c_{13}
 u_0^2\ln^2\frac{r}{r_{UH}}\Big],
\end{eqnarray}
where $m$ is an integral constant. Since $r^2\gg\ln r$ and $(s\cdot\chi)\sim\sqrt{\Lambda}r$ as $r\rightarrow\infty$, the cosmological constant would be non-negative. Therefore this solution can be asymptotically dS. It is easy to see that there are four independent constants $(m,u_0,r_{UH},Q)$, i.e., the universal horizon $r_{UH}$  cannot be showed via mass parameter $m$, which is unacceptable.

When $c_{13}=0$, it also reduces quasi-BTZ black hole
\begin{eqnarray}
  &&e(r)=-m-\Big[\Lambda r^2+(c_{14}u_0^2+\frac{Q^2}{2})\ln\frac{r}{l}\Big],
\end{eqnarray}
which is similar to (\ref{quasibtz}).

The surface gravity at the universal horizon can be computed and given by
\begin{eqnarray}\label{kappauh2}
 \kappa_{UH}=\frac{u_0}{2r_{UH}}
\sqrt{m+\frac{1}{1-c_{13}}\big[\Lambda r_{UH}^2+(\frac{Q^2}{2}+c_{14}u_0^2)\frac{r_{UH}}{l}\big]}.
\end{eqnarray}
From (\ref{SmarrFb}) and (\ref{tmassB}), one get the Komar integral
\begin{eqnarray}
0\cdot MG_{\ae}-\frac{\pi Q^2}{2}-\pi c_{14}u_0^2=2\pi r_{UH}[\Lambda r_{UH}+(1-c_{13})K_0(s\cdot\chi)|_{UH}].
\end{eqnarray}
However the relation between $m$ and $r_{UH}$ is unknown, so that we cannot derive the Smarr formula and the first law at universal horizon. In Ref. \cite{sotiriou}, the authors used Brown-Henneaux AdS boundary conditions to show that $c_{14}(\sim\eta)$ should be zero. So in the case of $c_{14}\neq0$, it is dS asymptotics, instead of AdS asymptotics. But neither its universal horizon nor black hole mass is under determined.

\section{Conclusions}
In this paper, we have studied  the Einstein-Maxwell-aether theory in 3-dimensions, and  found two new classes  of charged black hole solutions
for the special choices of the coupling constants: (1) $c_{14}=0,~c_{123}\neq0$, and (2) $c_{14}\neq0,~c_{123}=0$. There have three asymptotic forms, dS, flat or AdS, unlike general relativity, dependant on the bare(effective) cosmological constant $\Lambda(\bar\Lambda)$.

In the first case $c_{14}=0,~c_{123}\neq0$, the universal horizon depends on its electric charge $Q$, cosmological constant $\Lambda$ and aether constant $\Lambda'$. When $c_{13}\;  (\equiv c_1 + c_3)$ and $\Lambda'$ are very small and approach to zero, the solutions  reduce to the usual charged BTZ black hole solution but now with an universal horizon, which is very interesting. In the second case $c_{14}\neq0,~c_{123}=0$, the universal horizon may not exist though there have solutions with Killing horizon.

To study the solutions further, we have considered their surface gravity and constructed the Smarr formula at the universal horizons in three cases: (quasi)asymptotically flat $Q=0,\Lambda'^2=\Lambda/(c_2+c_{123})$, aether asymptotically flat $Q=0,\Lambda'=0$ and quasi-BTZ black hole spacetime $Q\neq0,r_{\ae}=0$. As for (quasi)asymptotically flat and uncharged black hole, the exact Smarr formula and the first law of black hole mechanics have been constructed with the knowledge of black hole chemistry. It is showed that the universal horizon may have an entropy and, the black hole mass is interpreted as an enthalpy of spacetime, the cosmological constant as a vacuum pressure. As for aether asymptotically flat and uncharged black hole, the slightly modified Smarr formula and the first law have been constructed. As for quasi-BTZ black hole, also the slightly modified Smarr formula and the first law have been constructed in the presence of electrical charge $Q$.

For the general cases  of nonzero $Q, \Lambda', r_{\ae}$, and $\Lambda'^2\neq\Lambda/(c_2+c_{123})$, instead of above three special cases, one cannot build the Smarr formula and first law. However, if one modifies the black hole mass via universal horizon temperature, or considers together with the couples between electromagnetical field and aether field, the first law would be constructed. Therefore, the universal horizon, as well as Killing horizon, can have an thermodynamical interpretation and, holography may be extended to it.

\begin{acknowledgments} C.D.  was supported by NNSFC No. 11247013, Hunan Provincial NSFC No. 2015JJ2085 and QSQC1203. C.L. was supported by special fund of NNSFC No. 11447168. A.W. was supported
in part by Ci\^{e}ncia Sem Fronteiras, No.
A045/2013 CAPES, Brazil and NNSFC No.
11375153, China. J.J. was supported by NNSFC No. 11475061; the SRFDP No. 20114306110003.
\end{acknowledgments}


\vspace*{0.2cm}
 \begin{thebibliography}{99}
 \baselineskip=0.6 cm


\bibitem{mattingly} D.~Mattingly,
Living Rev.\ Rel.\  {\bf 8}, 5 (2005).
\bibitem{bhattacharyya2} J. Bhattacharyya, Ph.D dessertation, University of New Hamspshire, Durham, NH, USA (2013).
\bibitem{shomer} A.~Shomer,  arXiv:0709.3555.
\bibitem{jacobson} T.~Jacobson and D.~Mattingly,
    Phys.\ Rev.\ D {\bf 64},   024028 (2001).
\bibitem{abdo}  A. A. Abdo {\it et al}., Nature {\bf462}, 331 (2009).
\bibitem{jacobson2} T. Jacobson and D. Mattingly, Phys. Rev. D {\bf 64}, 024028 (2001); T. Jacobson, Proc. Sci. QG-PH, {\bf 020} (2007) [arXiv:0801.1547].
\bibitem{bekenstein2004} J. D. Bekenstein, Phys. Rev. D {\bf70}, 083509 (2004), [Erratum: Phys. Rev.D {\bf71},069901(2005)]; T. G. Zlosnik, P. G. Ferreira, and G. D. Starkman, Phys. Rev.D {\bf74}, 044037 (2006); T. G. Zlosnik, P. G. Ferreira, and G. D. Starkman, Phys. Rev. D {\bf75}, 044017 (2007); L. Blanchet and S. Marsat, Phys. Rev. D {\bf84}, 044056 (2011); M. Bonetti and E. Barausse, Phys. Rev. D {\bf91}, 084053 (2015).
\bibitem{balakin2} A. B. Balakin and H. Dehnen, Phys. Lett. B {\bf 681}, 113 (2009).
\bibitem{jacobson3} T.~Jacobson and D.~Mattingly,
     Phys.\ Rev.\ D {\bf 63}, 041502 (2001).
\bibitem{jacobson4} T.~Jacobson and D.~Mattingly,
     Phys.\ Rev.\ D {\bf 70}, 024003 (2004).
\bibitem{blas2} D.~Blas and S.~Sibiryakov, Phys.\ Rev.\ D {\bf 84}, 124043 (2011).
\bibitem{lin} K. Lin, O. Goldoni, M. F. da Silva and A. Wang, Phys. Rev. D {\bf91},  024047 (2015).
\bibitem{UHs} K. Lin, F.-W. Shu, A. Wang, and Q. Wu, Phys. Rev. D {\bf91}, 044003 (2015);F.-W. Shu, K. Lin, A. Wang, and Q. Wu, JHEP {\bf 04}, 056 (2014);
                          K. Lin, E. Abdalla, R.-G.  Cai, and A. Wang,   Inter. J.  Mod.  Phys. D {\bf 23}, 1443004 (2014);
                          P. Horava, A. Mohd, C. M. Melby-Thompson, P. Shawhan, Gen. Rel. Grav. {\bf 46},   1720 (2014);
                          S. Janiszewski, A. Karch, B. Robinson, and D. Sommer, JHEP {\bf 04}, 163 (2014);
                         C. Eling and Y. Oz, JHEP {\bf 11}, 067 (2014);
                         M. Saravani, N. Afshordi, and R.B. Mann, Phys. Rev. D {\bf 89}, 084029 (2014);
                          A. Mohd, arXiv:1309.0907;
                          B. Cropp, S. Liberati, and M. Visser, Class. Quantum Grav. {\bf 30}, 125001 (2013).

\bibitem{barausse} E. Barausse, T. Jacobson and T. P. Sotiriou, Phys. Rev. D {\bf83}, 124043 (2011).
\bibitem{latta} J. Latta and G. Leon, arXiv: 1606.08586.
\bibitem{balakin2016} A. B. Balakin, Phys. Rev. D  {\bf94}, 024021 (2016).
\bibitem{gurses} M. G\"{u}rses and \c{C}. \c{S}ent\"{u}rk, arXiv: 1604.02266.
\bibitem{cropp2016}B. Cropp, S. Liberati and R. Turcati, Phys. Rev. D {\bf94}, 063003 (2016).
\bibitem{berglund} P. Berglund, J. Bhattacharyya and D. Mattingly, Phys. Rev. D {\bf85}, 124019 (2012).
\bibitem{bhattacharyya} J. Bhattacharyya and D. Mattingly, Int. J. Mod. Phys. D {\bf 23},  1443005 (2014).
\bibitem{ding} C. Ding, A. Wang and X. Wang, Phys. Rev. D {\bf 92}, 084055 (2015).
\bibitem{barausse2016} E. Barausse, T. P. Sotiriou and I. Vega, Phys. Rev. D {\bf 93}, 044044 (2016).
\bibitem{banados} M. Ban\~ados, C. Teitelboim and J. Zanelli, Phys. Rev. Lett. {\bf69}, 1849 (1992); M. Ban\~ados, M. Henneaux, C. Teitelboim and J. Zanelli, Phys. Rev. D {\bf48}, 1506 (1993).
\bibitem{donnay} L. Donnay, arXiv: 1602.09021.
\bibitem{sotiriou} T. Sotiriou, I. Vega, and D. Vernieri, Phys. Rev. D {\bf 90}, 044046 (2014).
\bibitem{eling06} C. Eling and T. Jacobson, Phys. Rev. D {\bf74},  084027 (2006).
\bibitem{basu} S. Basu, J. Bhattacharyya, D. Mattingly and M. Roberson,     Phys. Rev. D {\bf93},  064072 (2016).
\bibitem{frassino} A. M. Frassino, R. B. Mann and J. R. Mureika, Phys. Rev. D {\bf92}, 124069 (2015).
\bibitem{eling} C. Eling, Phys. Rev. D {\bf73},  084026 (2006).

\bibitem{garfinkle} D.~Garfinkle, T.~Jacobson,
  Phys. Rev. Lett. {\bf107},  191102 (2011).
\bibitem{balakin}A. B. Balakin and J. P. S. Lemos, Ann. Phys. {\bf350}, 454 (2014); S. Janiszewski, A. Karch, B. Robinson, and D. Sommer, JHEP {\bf 04}, 163 (2014).

\bibitem{kastor} D. Kastor, S. Ray and J. Traschen, Class. Quantum
Gravi. {\bf26}, 195011 (2009).
\bibitem{kubiznak}  D. Kubiznak and R. B. Mann, Can. J. Phys. {\bf93}, 999 (2015).
 \bibitem{BCH}     J. M. Bardeen, B. Carter, and S. W. Hawking, Commun. Math. Phys. {\bf 31}, 161 (1973).
\bibitem{jacobson5}
  T. Jacobson, in CPT and Lorentz Symmetry: Proceedings
of the Fourth Meeting, Bloomington, USA, 11 August
2007, edited by V. A. Kostelecky¡ä (World Scientific,
Singapore, 2008), p. 92, arXiv:0711.3822.
\bibitem{cropp} B. Cropp, S. Liberati and A. Mohd, Phys. Rev. D {\bf89},  064061 (2014).
\bibitem{ding2} C. Ding, A. Wang, X. Wang and T. Zhu, Nucl. Phys. B {\bf 913}, 694 (2016).








\end{thebibliography}
\end{document}